\documentclass{ieeeaccess}
\usepackage{cite}
\usepackage{amsmath,amssymb,amsfonts}
\usepackage{algorithmic}
\usepackage{graphicx}
\usepackage{textcomp}
\usepackage{siunitx}
\usepackage{pifont}
\usepackage{makecell}
\usepackage{multirow}

\usepackage{footmisc}
\usepackage{hyperref}

\def\BibTeX{{\rm B\kern-.05em{\sc i\kern-.025em b}\kern-.08em
    T\kern-.1667em\lower.7ex\hbox{E}\kern-.125emX}}
\usepackage{booktabs, makecell}

\usepackage{xspace}
\makeatletter
\newcommand*{\etal}{%
    \@ifnextchar{.}%
        {\textit{et al}}%
        {\textit{et al.}\@\xspace}%
}
\makeatother

\ifCLASSOPTIONcompsoc
    \usepackage[caption=false, font=normalsize, labelfont=sf, textfont=sf]{subfig}
\else
\usepackage[caption=false, font=footnotesize]{subfig}
\fi

\begin{document}

\doi{xx.xxxx/ACCESS.xxx.DOI}

\title{Improving Post-processing of Audio Event Detectors using Reinforcement Learning}

\author{\uppercase{Petros Giannakopoulos}\authorrefmark{1}, 
\uppercase{Aggelos Pikrakis}\authorrefmark{2}, \IEEEmembership{Member, IEEE}, \uppercase{and Yannis Cotronis}\authorrefmark{3}}
\address[1]{National and Kapodistrian University of Athens (e-mail: petrosgk@di.uoa.gr)}
\address[2]{University of Piraeus (e-mail: pikrakis@unipi.gr)}
\address[3]{National and Kapodistrian University of Athens (e-mail: cotronis@di.uoa.gr)}
\tfootnote{Research in this paper was supported by the Hellenic Foundation for Research and Innovation (H.F.R.I.) under the ``1st Call for H.F.R.I. Research Projects to support Faculty Members \& Researchers and the Procurement of High-Cost Research Equipment Grant'' (Project Number: 3449).}

\markboth
{Petros Giannakopoulos \headeretal: Improving Post-processing of Audio Event Detectors using Reinforcement Learning}
{Petros Giannakopoulos \headeretal: Improving Post-processing of Audio Event Detectors usning Reinforcement Learning}

\corresp{Corresponding author: Petros Giannakopoulos (e-mail: petrosgk@di.uoa.gr)}

\begin{abstract}
We apply post-processing to the class probability distribution outputs of audio event classification models and employ reinforcement learning to jointly discover the optimal parameters for various stages of a post-processing stack, such as the classification thresholds and the kernel sizes of median filtering algorithms used to smooth out model predictions. To achieve this we define a reinforcement learning environment where: 1) a state is the class probability distribution provided by the model for a given audio sample, 2) an action is the choice of a candidate optimal value for each parameter of the post-processing stack, 3) the reward is based on the classification accuracy metric we aim to optimize, which is the audio event-based macro F1-score in our case. We apply our post-processing to the class probability distribution outputs of two audio event classification models submitted to the DCASE Task4 2020 challenge. We find that by using reinforcement learning to discover the optimal per-class parameters for the post-processing stack that is applied to the outputs of audio event classification models, we can improve the audio event-based macro F1-score (the main metric used in the DCASE challenge to compare audio event classification accuracy) by 4-5\% compared to using the same post-processing stack with manually tuned parameters.
\end{abstract}


\titlepgskip=-15pt

\maketitle

\begin{figure*}
\centerline{\includegraphics[width=\columnwidth]{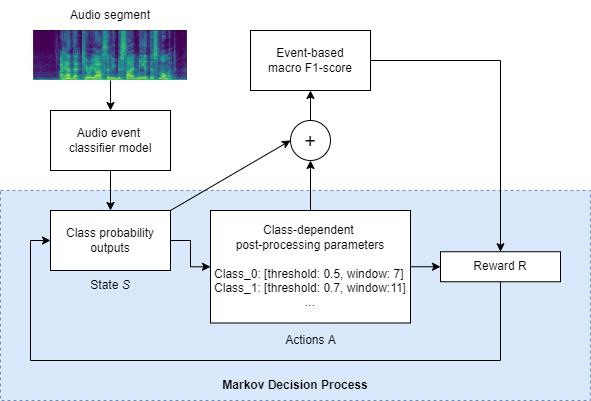}}
\caption{Illustration of the proposed MDP approach to optimize the parameters of a post-processing stack for an audio event classification model. The RL agent receives the \textit{class probability outputs} of the model and chooses the parameters per class for each part of the post-processing stack (in this case thresholding and median filtering). The post-processing stage with the given parameters is subsequently applied to the model outputs and the F1-score is updated. The agent then receives a reward based on the achieved score and proceeds to the next audio segment.}
\label{fig:mdp}
\end{figure*}

\section{Introduction}

\IEEEPARstart
As it is common with most audio analysis methods that operate by taking classification decisions at the audio frame level, the goal of a final post-processing stage is to increase the  classification performance of an inevitably imperfect model by smoothing out the sequence of predictions, removing classification noise like false-positive or false-negative decisions and improving the boundaries of detected audio events. A typical post-processing stage consists of a soft output thresholding step followed by a smoothing operation, with median filtering being a popular choice. The parameters of such operations are usually hand-crafted following some grid search technique. It can therefore be understood that post-processing schemes have received significantly less attention over recent years, compared to the core audio classification algorithms. To remove the need for such manual parameter tuning, we propose a reinforcement learning \cite{sutton2018reinforcement} approach that automatically computes parameter values of the post-processing pipeline operations by treating them as learnable parameters that can be applied to the class probability distribution outputs of audio event detection and classification models. A widely accepted problem formulation for common audio event detection and classification problems can be found in the DCASE Task4 challenge task descriptions, where, for example, according to the  sound event detection task specification (Task4 in recent years) \cite{Turpault2019_DCASE}, a model must be able to identify the endpoints of each audio event in a given audio recording along with the class label of each audio event. Examples of such audio events are human speech, dog barks, vacuum cleaner sounds and so on. In general, audio events may overlap and multiple classes are expected to be accommodated by an algorithm ($10$ classes is a typical scenario). Following the DCASE Task4 challenge practice, audio event detection algorithms are primarily evaluated based on the so-called event-based macro F1-score \cite{app6060162}, according to which a certain tolerance is permitted around the detected event endpoints, like a $200$ ms tolerance collar on event onsets and a $200$ ms (or 20\% of the events length tolerance) collar on event offsets. In order to validate our approach, we apply the proposed RL-based post-processing method to the class probability outputs of two audio event detection models that were submitted to Task 4 of the DCASE-2020 challenge: a) A baseline detector which was developed by the organizers and achieved a $40.1\%$ event-based macro-F1 score \cite{Turpault2019_DCASE} on the evaluation set of the challenge and b) The best performing detector of the 2020 challenge, ConformerSED \cite{miyazaki2020conformer} which achieved a $47.7\%$ F1-score on the same evaluation set. Both systems apply class probability thresholding and a subsequent median filtering step at a post-processing stage and are, therefore, good candidates for our method. In both challenge submissions, the respective parameters were manually tuned by the method proposers using a coarse grid search in the parameter space and were class-independent, i.e., the same values were applied to all classes. If, instead, our RL-based method is used  to automatically fine-tune a classification threshold and median window length per audio event class, the F1-score of the baseline method increases to $44.5\%$, and the F1-score of the best performing method increases to $52.4\%$. We therefore observe a $4\%$ to $5\%$ performance improvement on average, over very different methods, if the parameters of a standard post-processing pipeline are tuned using RL, assuming that the RL agent has access to the soft outputs of a method for a set of training samples.

Our contribution can be summarised as follows:
\begin{itemize}
    \item We introduce a process based on reinforcement learning for optimizing the parameters of the typical post-processing stack (thresholding and median filtering) applied to the prediction outputs of an audio event detection system.
    \item We demonstrate that this reinforcement learning optimization approach of the post-processing parameters provides superior audio event detection performance compared to the typical grid search optimization approaches.
\end{itemize}

The code for this work is available on GitHub \footnote{\url{https://github.com/petrosgk/SED_RL_post_processing}}.

\section{Related Work}

The work in \cite{cances2019evaluation} conducted an evaluation of post-processing algorithms that were applied on the output of the baseline DCASE-2018 Task4 audio event detector model and another submitted model. They concluded that post-processing has a large impact on the final classification performance as measured by the event-based F1-score and that post-processing algorithms performed better when they were class-dependent. The authors in \cite{xia2017frame} proposed a frame-wise dynamic threshold selection method for polyphonic sound event detection and showed that it was superior to using a fixed threshold for all audio frames. To estimate the threshold per frame, they used contour-based and regressor-based methods. Our method adds to the post-processing approaches outlined in these previous works by introducing RL as a way of optimizing the parameters of post-processing, as far as both thresholding and median filtering are concerned. Overall, however, the optimization of the post-processing stage of audio event detection systems has not received enough attention in literature so far, in our opinion, while it has been shown that it can have a significant impact on the final performance of audio event detection systems \cite{cances2019evaluation}. Regarding RL methods for audio analysis tasks, recent systems primarily address the application of autonomous agents that are capable of leveraging auditory information to perform diverse tasks given an environment,  e.g., \cite{woubie2019autonomous, hegde2021agents, lathuiliere2018deep, giannakopoulos2021deep, van2022deep}, but, to the best of our knowledge, our work is the first to formulate parameter tuning at the post-processing stage as a RL task and employ a RL algorithm to automatically compute parameter values for improving the final classification performance of audio event detector systems.

\section{Method description}
\subsection{Post-processing fundamentals for Audio Event Detection Methods}
Given an audio recording, audio event detectors usually perform  \textit{strong labeling} or \textit{weak labeling} tasks. The former refers to problems where both endpoints and class labels of audio events  need to be reported, whereas the latter refers to tasks where only the presence or absence of an audio event type needs to be returned, thus skipping the need for endpoint detection. Note that it is common to assume that in a multi-class, strong labelling context, two or more events may overlap. As it is usually the case, a soft decision output is computed per class over the duration of the recording and a threshold is subsequently applied for each event class. Soft values above the threshold indicate class presence in the respective frames. For the \textit{strong labeling} task, when successive frames yield values above a threshold, they are grouped to form an event. In the case of \textit{weak labeling} problems, thresholding is used to identify which event classes are  present in the audio segment by collecting a percentage of frames that pass the threshold test per class.

\begin{figure}
\centerline{\includegraphics[width=\columnwidth]{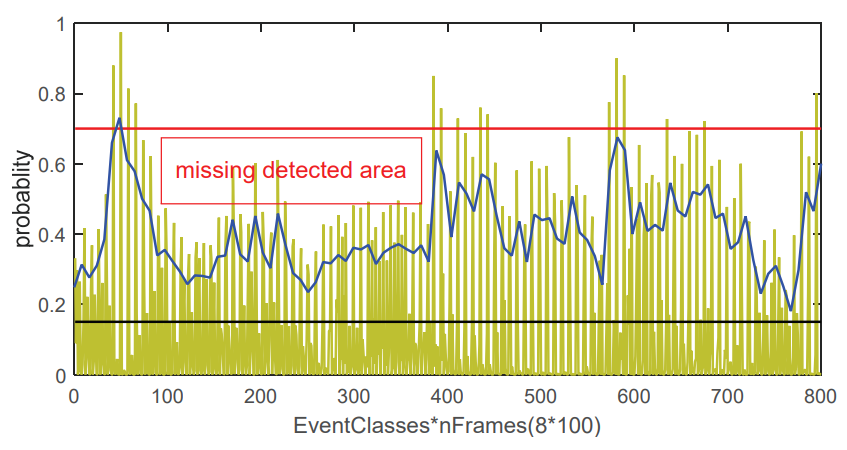}}
\caption{Strong (frame-level) labeling of audio events at the output of an AED system, taken from \cite{xia2017frame}. Global fixed thresholds can lead to missed events and wrong classifications.}
\label{fig:missed_threshold}
\end{figure}

As an example, if a detector model outputs posterior class probabilities at the frame level or at the audio segment level, a threshold of $0.5$ means that an audio event is considered to be present if the respective class probability is higher than $0.5$. A straightforward threshold selection approach has been to set a fixed threshold value for all classes. This type of naive threshold selection may work on certain occasions but it will fail to provide good results for many distributions of model outputs over the audio event classes (Figure \ref{fig:missed_threshold}) \cite{xia2017frame}. We can also set different thresholds per event class, which can potentially take into account variations in the model's behaviour for different event types. Apart from setting thresholds on the dataset level, we can increase threshold selection granularity and set thresholds  at the audio clip level, or, in the case of \textit{strong labels}, even at frame level. Our work computes a  threshold per class at the audio clip level and it is suitable for both \textit{strong labeling} and \textit{weak labeling} tasks.

\begin{figure}
\centerline{\includegraphics[width=\columnwidth]{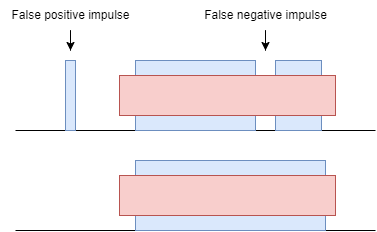}}
\caption{Real (red) vs. predicted (blue) boundaries of an audio event at the frame level, before (up) and after (down) applying median filtering. Median filtering can be used to remove impulsive false positive or false negative detections of frames (hills and valleys, respectively), as being part or not of an audio event, by smoothing out the detected audio event boundaries.}
\label{fig:med_filt}
\end{figure}

Median filtering is a second post-processing step that is used especially in \textit{strong labeling} tasks to improve event detection accuracy. It can remove certain artifacts of the event detector, like false positive or false negative decisions that only last a few successive frames and can therefore smooth out the detected audio event boundaries (Figure \ref{fig:med_filt}) and reduce over-segmentation. The median window size can play an important role and different  sizes may be appropriate for various types of events \cite{cances2019evaluation}, in order to balance undesirable (excessive) smoothing of event boundaries and decision noise suppression.

\subsection{Post-processing Optimization using Reinforcement Learning}
Based on the above description, a standard post-processing pipeline of an Audio Event Detector (AED) system performing \textit{strong labeling} consists of thresholding and, optionally, median filtering. The problem is to estimate threshold values that strike a balance between precision and recall for each class, as well as median window sizes that remove detection artifacts without causing excessive smoothing while taking into account the characteristics of the system response for each audio event class.

Our method assumes that there exists a set of reasonable values for each parameter of the post-processing pipeline, e.g. a  $(0.0, 1.0)$ continuous range for the threshold selection step and discrete $[3, 5, 7,\ldots,21]$ values for the median filtering window size. Contrary to the range of threshold values, the range of median filtering window sizes is unbounded. For this reason, we had to select a range of median filtering window sizes that are sensible and most commonly encountered in AED systems literature. Our goal then becomes to find ``optimal'', class-dependent parameters for each audio segment in the dataset. We can therefore consider the problem from a MDP perspective, i.e., as solving a MDP (Figure \ref{fig:mdp}), where a state $S$ is the AED output for a given audio segment, i.e., a matrix of class probabilities, one row per frame, like in (Figure \ref{fig:missed_threshold}). Given $S$, we select a threshold value and a median window length from the respective sets of possible values. If we are dealing with a class-dependent optimization scenario, two different values will be selected per class. Therefore, the action vector, $A$, of the RL agent, will consist of $2\times N$ elements, where $N$ is the number of classes. If we are interested in a simpler, class independent setting, there will only be two values per audio clip and the action vector of the RL agent will therefore be two-dimensional. Then the processing stack with the selected parameters is applied to the AED outputs and the event-based macro F1-score for that audio segment is evaluated. The higher the F1-score, the higher the reward, $R$, of taking action $A$ in state $S$. The goal is to maximize the average reward  over the entire dataset, which directly correlates to maximizing the average F1-score. Given that the F1-score is in the range $[0\%, 100\%]$, and the reward should be normalized in the range $[0.0, 1.0]$, for training stability purposes, it follows that the relationship between the reward $R$ and the F1-score would simply be $R = F1 / 100$.

\begin{figure}
\centerline{\includegraphics[width=\columnwidth]{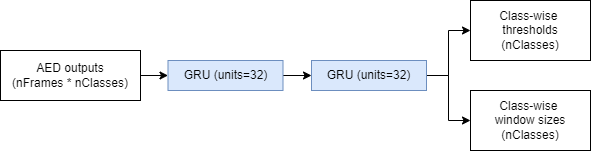}}
\caption{Neural network architecture used as policy function approximator. It receives as input the AED frame-level event class probabilities (the state $S$) and the outputs are vectors of length $nClasses$ for the class-dependent thresholds and window sizes, or scalars if class-independent.}
\label{fig:net_arch}
\end{figure}

\begin{table}[t]
\resizebox{\columnwidth}{!}{%
\begin{tabular}{@{}lr@{}}
\toprule
\textbf{Parameter}                   & \textbf{Value}                           \\ \midrule
sample rate                          & 16000                                    \\
max audio segment length (seconds)   & 10                                       \\
number of classes                    & 10                                       \\
threshold range                      & (0.0, 1.0)                               \\
set of median filtering window sizes & {[}3, 5, 7, 9, 11, 13, 15, 17, 19, 21{]} \\ \bottomrule
\end{tabular}%
}
\caption{Values of parameters specific to the DCASE Task4 dataset. Used for defining the reinforcement learning environment the PG algorithm will be trained on.}
\label{tab:dcase_dataset_params}
\end{table}

\begin{table}[t]
\centering
\resizebox{2in}{!}{%
\begin{tabular}{@{}lr@{}}
\toprule
\textbf{Parameter}     & \textbf{Value}   \\ \midrule
batch size             & 4                \\
memory size            & 10000 iterations \\
optimizer              & Adam             \\
learning rate          & 0.001            \\
update frequency       & 4                \\
discount factor ($\gamma$)    & 0.99             \\
entropy regularization & 0.001            \\
GRU state size         & 32               \\ \bottomrule
\end{tabular}%
}
\caption{Values of hyperparameters used during training of the Policy Gradient algorithm.}
\label{tab:pg_training_params}
\end{table}

\begin{table*}[]
\centering
\begin{tabular}{@{}l|cc|cc@{}}
\toprule
                                  & \multicolumn{2}{c|}{\textbf{Baseline}}                           & \multicolumn{2}{c}{\textbf{ConformerSED}}                        \\ \midrule
\textbf{Post-processing methods}  & \textbf{Thresholding} & \textbf{Thresholding + Median Filtering} & \textbf{Thresholding} & \textbf{Thresholding + Median Filtering} \\ \midrule
                                  & Event-based F1 (\%)   & Event-based F1 (\%)                      & Event-based F1 (\%)   & Event-based F1 (\%)                      \\
Original                          & 38.5                  & 40.1                                     & 44.7                  & 47.7                                     \\
Class-dependent PG optimization   & 42.4                  & \textbf{44.5}                            & 49.2                  & \textbf{52.4}                            \\
Class-independent PG optimization & 41.3                  & 43.4                                     & 48.0                  & 51.1                                     \\ \bottomrule
\end{tabular}
\caption{Event-based macro F1-score for both Baseline and ConformerSED systems, when the original parameters for thresholding and median filtering window are used, and when optimized via Policy Gradient reinforcement learning algorithm. We try both class-dependent and class-independent optimization. We also compare the performance when only thresholding is used and when both thresholding and median filtering post-processing are used. The results are obtained on the validation set provided by the DCASE Task4 2020 challenge.}
\label{tab:pg_opt_results}
\end{table*}

\subsection{Policy Gradient Optimization}

To solve the MDP we employ Policy Gradient (PG) reinforcement learning method \cite{sutton1999policy, schaarschmidt2018lift}, where the policy function approximator is a simple recurrent neural network made up from a stack of 2 GRU layers \cite{cho2014properties} (Figure \ref{fig:net_arch}). We chose a recurrent architecture in order to be able to better capture time-dependent information found in the sequences of predictions of an AED system, given that the AED system itself usually performs classification over audio signal feature sequences that it receives as input. The objective of the PG algorithm is to find a policy function $\pi_\theta$, where $\theta$ are the weights of the neural network, which adequately approximates the optimal policy $\pi^*$. When $\pi^*$ is followed it yields the maximum achievable mean reward $R$ over the dataset, which translates to the maximum achievable event-based macro F1-score.

The PG algorithm iteratively makes $\pi_\theta$ converge to $\pi^*$ by assigning a value to each state $S$, which represents the average reward that it expects to get in the future by following the current policy  $\pi_\theta$ in $S$, and it is directly related to the average expected F1-score. More formally:
\begin{equation}
    V^{\pi}(S) = \mathbb{E}[\sum_{k=0}^\infty\gamma^kr_{t+k}]
\end{equation}

where $t \in [0,T]$, with $T = len(dataset)$, is the iteration index, $r_t$ is the reward at iteration $t$, and $\gamma \in (0, 1]$ is the future reward discount factor, which governs how much "weight" the algorithm should place on long-term vs short-term rewards.

After the value of state $S$ is calculated, the next step is to estimate the value (or advantage) of taking action $A$ in $S$ at iteration $t$ as the sum of the expected values of future states, resulting from taking action $A$ in state $S$, denoted here as $s_t$ and $a_t$:
\begin{equation}
    A_t(a_t, s_t) = \delta_t + \gamma\delta_{t+1} + \dots + \gamma^{T-t+1}\delta_{T-1},
\end{equation}

where
\begin{equation}
    \delta_t = r_t + \gamma V^{\pi}(s_{t+1}) - V^{\pi}(s_t),
\end{equation}

At each iteration $t$ the policy approximator network outputs log-probabilities (after the application of log-softmax) for each action, multipled by the estimated advantage of that action:
\begin{equation}
    \log{\pi_\theta(a_t | s_t)}
\end{equation}

The network outputs two actions $a_t$ given state $s_t$. In the case of class-dependent optimization each $a_t$ is a vector of log-probabilities, of length $nClasses$. Each vector entry is another vector of length $nValues$, where $nValues$ is the number of possible values for the threshold and for the median filtering window size. In the class-independent case $a_t$ is simply a vector of length $nValues$ containing scalars. Since threshold values are continuous, they are sampled from a uniform distribution in the range $(0.0, 1.0)$.

After multiplying the action log-probabilities (4) with the advantage estimate for each action (2), the final objective optimized by PG for a batch of log-probabilities is:
\begin{equation}
    L^{PG}(\theta) = \hat{\mathbb{E}}[\log{\pi_\theta(a_t | s_t)}\hat{A_t}]
\end{equation}

where $\mathbb{E}$ is the mean over the batch. To promote exploration of the state space and avoid convergence to a sub-optimal policy, an entropy regularization term $H(\theta) = -\pi_\theta\log{\pi_\theta}$ is added to the PG objective \cite{mnih2016asynchronous} so that the final objective becomes:
\begin{equation}
    L = L^{PG}(\theta) + \beta H(\theta)
\end{equation}

where $\beta$ is the weight of the entropy regularization term in the final objective. In place, or in addition to entropy regularization, other methods for facilitating exploration of the state space could be used, such as the addition of gaussian or uniform noise to the continuous threshold predictions of the network or randomly replacing the discrete median filtering window size predictions of the network with other values sampled from the set of window sizes.

\section{Experiments}

Similar to \cite{cances2019evaluation}, we apply our method to the optimization of the post-processing stack of two AED systems: a baseline approach and a state-of-the-art method, to cover the two extremes of the performance spectrum. Table \ref{tab:dcase_dataset_params} shows the values of the parameters that define the state and action spaces of the reinforcement learning environment on which the Policy Gradient algorithm is trained. Specifically, the state space is defined by the \textit{sample rate}, \textit{maximum length of audio segments} and \textit{number of classes} present in the dataset. The sets of possible \textit{threshold values} and \textit{median filter window sizes} define the action space. Table \ref{tab:pg_training_params} shows the values of the Policy Gradient algorithm hyperparameters that are used for training. Specifically, the choice of \textit{batch size} was constrained by the available system memory, or how many copies of the dataset could fit in memory. \textit{Memory size} refers to the history of training iterations, or interactions of the RL agent with the MDP environment, that are buffered and is also constrained by available system memory. The rest of the parameters, such as the choice of \textit{optimizer}, \textit{learning rate}, \textit{update frequency} were taken from \cite{mnih2016asynchronous}. The training parameters remained identical for both AED systems that the PG algorithm was trained on. The most important ones are the \textit{discount factor} ($\gamma$) and the \textit{entropy regularization} weight. A discount factor close to $1$ is needed to assign more importance to rewards gained in the future, because the F1-score is only calculated once at the end of the dataset after the predictions for all audio segments have been accumulated. The entropy regularization weight value achieves a balance between exploration and exploitation: a low value leads to faster convergence but less exploration of the state space, which in turn can lead to convergence to a sub-optimal F1-score.

The results of our experiments are grouped in Table \ref{tab:pg_opt_results}, for the two AED systems that we evaluated. The first line of the table shows the event-based macro F1-score when using the same parameters as the authors of the respective work. The second line shows the F1-score that was achieved by optimizing the threshold value and median filter size per class (class-dependent) using policy gradient reinforcement learning. Finally, the third line shows the class-independent PG optimization, where only global threshold and median filter window values are optimized and used for all classes. We also test the effect of median filtering on the AED performance by removing it from the post-processing stack to only use thresholding.

For the Baseline AED system, the authors use a fixed event class probability threshold of $0.5$ and a fixed window size of $7$ for the median filter. Applying median filtering after thresholding improves F1-score by $1.6\%$, from $38.5\%$ to $40.1\%$. Class-dependent optimization of thresholding and median filtering parameters via RL leads to a F1-score improvement of $4.4\%$ over the original fixed values, from $40.1\%$ to $44.5\%$. Class-independent optimization gives a slightly lower improvement of $3.3\%$ over the original values.

For the ConformerSED system, the authors use a fixed threshold of $0.5$ but perform a coarse grid search separately for each event class to tune the median filter length.
Class-dependent RL optimization computes on a class basis better values for the median filter length, as the F1-score improves from $47.7\%$ to $52.4\%$, an increase of $4.7\%$. Again, class-independent optimization performs slightly worse than the class-dependent one, with an improvement of $3.4\%$ over the original values.

Overall, a reinforcement agent that earns to compute ``optimal'' values for class thresholds and median filter sizes using policy gradient optimization, yields performance improvements for both AED methods under study. It can also be observed that class-dependent optimization yields better results compared to the class-independent (global) optimization case.

As PG optimizes for a reward-based objective, we can use it to indirectly optimize a non-differentiable objective, such as the F1-score in our case, by translating increases or decreases of that objective to positive or negative credit assignment. Since the F1-score is in the $[0\%, 100\%]$ range of values, we simply do a straightforward linear mapping to a reward in the $[0.0, 1.0]$ range.

\section{Limitations}

The computational time that is required to learn the policy function with a neural network approximator can be non-trivial, depending on the number of audio segments. Given that each audio segment is treated as a state of the MDP, the MDP will grow larger with the amount of audio segments. This in turn will increase the time required for the PG algorithm to reach convergence. In a standard experimental setup with low-cost hardware (Intel Core i9-10900K processor, 32GB RAM) convergence was achieved after approximately $2$ hours of processing time on the DCASE-Task4 training dataset, which consists of approximately $2500$ audio segments. However, the computational time required is, in our opinion, offset by the superior final classification accuracy achieved by our method. As a final remark, note that our method needs to be trained separately on each AED system output, i.e., we do not provide a global RL agent for all possible AED methods.

\section{Conclusion}

In this work we introduced reinforcement learning, and in particular policy gradient optimization, as a way of discovering optimal parameter values for the two most commonly used stages of the post-processing stack of an audio event detector: thresholding and median filtering. To that end, we reformulate this parameter optimization problem as a Markov Decision Process, where: 1) the state $S$ consists of the class probability outputs of the AED system for a given audio segment of the dataset, 2) the actions $A$ are the values selected as optimal by the RL algorithm for the thresholding and median filtering parameters, either on a per-class basis or globally, 3) the reward $R$ is the event-based macro F1-score achieved by the RL algorithm over the entire dataset using the selected parameter values for the post-processing stack. 

We tested our method on two AED systems submitted to the DCASE Task4 (Sound Event Detection in Domestic Environments) 2020 challenge: 1) A baseline AED system provided by the challenge organizers, 2) The system (ConformerSED) that achieved top ranking on the challenge, based on achieved event-based macro-F1 score. Both systems use a \textit{thresholding + median filtering} post-processing stack. We achieved an improvement of about $4\%$ to $5\%$ on average on the event-based macro-F1 score compared to the performance of these systems with their original parameterization of the post-processing stack.

A drawback of the proposed method is the requirement to train the PG algorithm separately on the outputs of each AED system. The computational requirements highly depend on the size of the dataset of audio segments. Another important prerequisite for the success of the method is the sufficient exploration of the state space to avoid converging on sub-optimal post-processing parameter values, generally known as exploration vs. exploitation dilemma in RL literature. However, when properly tuned, it can provide superior final classification performance for an AED system compared to performing exhaustive grid search for finding the optimal parameter values for the post-processing stack, especially on larger datasets. An interesting avenue of future research would be to simultaneously train the PG algorithm on the predictions of several AED systems for a particular dataset and assess its ability to generalize to other AED systems that it was not trained on. Another possibility for future work would be to try other, more sophisticated than Policy Gradient, reinforcement learning algorithms, with our method, such as Proximal Policy Optimization (PPO) \cite{schulman2017proximal} and Soft Actor-Critic (SAC) \cite{haarnoja2018soft}.

\bibliographystyle{IEEEtran}
\bibliography{main}

\begin{thebibliography}{10}
\providecommand{\url}[1]{#1}
\csname url@samestyle\endcsname
\providecommand{\newblock}{\relax}
\providecommand{\bibinfo}[2]{#2}
\providecommand{\BIBentrySTDinterwordspacing}{\spaceskip=0pt\relax}
\providecommand{\BIBentryALTinterwordstretchfactor}{4}
\providecommand{\BIBentryALTinterwordspacing}{\spaceskip=\fontdimen2\font plus
\BIBentryALTinterwordstretchfactor\fontdimen3\font minus
  \fontdimen4\font\relax}
\providecommand{\BIBforeignlanguage}[2]{{%
\expandafter\ifx\csname l@#1\endcsname\relax
\typeout{** WARNING: IEEEtran.bst: No hyphenation pattern has been}%
\typeout{** loaded for the language `#1'. Using the pattern for}%
\typeout{** the default language instead.}%
\else
\language=\csname l@#1\endcsname
\fi
#2}}
\providecommand{\BIBdecl}{\relax}
\BIBdecl

\bibitem{sutton2018reinforcement}
R.~S. Sutton and A.~G. Barto, \emph{Reinforcement learning: An
  introduction}.\hskip 1em plus 0.5em minus 0.4em\relax MIT press, 2018.

\bibitem{Turpault2019_DCASE}
N.~Turpault, R.~Serizel, A.~Parag~Shah, and J.~Salamon, ``{Sound event
  detection in domestic environments with weakly labeled data and soundscape
  synthesis},'' in \emph{{Workshop on Detection and Classification of Acoustic
  Scenes and Events}}, New York City, United States, October 2019.

\bibitem{app6060162}
A.~Mesaros, T.~Heittola, and T.~Virtanen, ``Metrics for polyphonic sound event
  detection,'' \emph{Applied Sciences}, vol.~6, no.~6, 2016.

\bibitem{miyazaki2020conformer}
K.~Miyazaki, T.~Komatsu, T.~Hayashi, S.~Watanabe, T.~Toda, and K.~Takeda,
  ``Conformer-based sound event detection with semi-supervised learning and
  data augmentation,'' in \emph{Detection and Classification of Acoustic Scenes
  and Events}, 2020, pp. 100--104.

\bibitem{cances2019evaluation}
L.~Cances, P.~Guyot, and T.~Pellegrini, ``Evaluation of post-processing
  algorithms for polyphonic sound event detection,'' in \emph{IEEE Workshop on
  Applications of Signal Processing to Audio and Acoustics (WASPAA)}.\hskip 1em
  plus 0.5em minus 0.4em\relax IEEE, 2019, pp. 318--322.

\bibitem{xia2017frame}
X.~Xia, R.~Togneri, F.~Sohel, and D.~Huang, ``Frame-wise dynamic threshold
  based polyphonic acoustic event detection,'' in \emph{INTERSPEECH}, 2017, pp.
  474--478.

\bibitem{woubie2019autonomous}
A.~Woubie, A.~Kanervisto, J.~Karttunen, and V.~Hautamaki, ``Do autonomous
  agents benefit from hearing?'' \emph{arXiv preprint arXiv:1905.04192}, 2019.

\bibitem{hegde2021agents}
S.~Hegde, A.~Kanervisto, and A.~Petrenko, ``Agents that listen: High-throughput
  reinforcement learning with multiple sensory systems,'' in \emph{IEEE
  Conference on Games (CoG)}.\hskip 1em plus 0.5em minus 0.4em\relax IEEE,
  2021, pp. 1--5.

\bibitem{lathuiliere2018deep}
S.~Lathuili{\`e}re, B.~Mass{\'e}, P.~Mesejo, and R.~Horaud, ``Deep
  reinforcement learning for audio-visual gaze control,'' in \emph{IEEE/RSJ
  International Conference on Intelligent Robots and Systems (IROS)}.\hskip 1em
  plus 0.5em minus 0.4em\relax IEEE, 2018, pp. 1555--1562.

\bibitem{giannakopoulos2021deep}
P.~Giannakopoulos, A.~Pikrakis, and Y.~Cotronis, ``A deep reinforcement
  learning approach to audio-based navigation in a multi-speaker environment,''
  in \emph{IEEE International Conference on Acoustics, Speech and Signal
  Processing (ICASSP)}.\hskip 1em plus 0.5em minus 0.4em\relax IEEE, 2021, pp.
  3475--3479.

\bibitem{van2022deep}
T.~Van~Nguyen, X.~Dai, I.~Khan, R.~Thawonmas, and H.~V. Pham, ``A deep
  reinforcement learning blind ai in darefightingice,'' \emph{arXiv preprint
  arXiv:2205.07444}, 2022.

\bibitem{sutton1999policy}
R.~S. Sutton, D.~McAllester, S.~Singh, and Y.~Mansour, ``Policy gradient
  methods for reinforcement learning with function approximation,''
  \emph{Advances in neural information processing systems}, vol.~12, 1999.

\bibitem{schaarschmidt2018lift}
M.~Schaarschmidt, A.~Kuhnle, B.~Ellis, K.~Fricke, F.~Gessert, and E.~Yoneki,
  ``Lift: Reinforcement learning in computer systems by learning from
  demonstrations,'' \emph{arXiv preprint arXiv:1808.07903}, 2018.

\bibitem{cho2014properties}
K.~Cho, B.~Van~Merri{\"e}nboer, D.~Bahdanau, and Y.~Bengio, ``On the properties
  of neural machine translation: Encoder-decoder approaches,'' in \emph{Eighth
  Workshop on Syntax, Semantics and Structure in Statistical Translation
  (SSST-8)}, 2014.

\bibitem{mnih2016asynchronous}
V.~Mnih, A.~P. Badia, M.~Mirza, A.~Graves, T.~Lillicrap, T.~Harley, D.~Silver,
  and K.~Kavukcuoglu, ``Asynchronous methods for deep reinforcement learning,''
  in \emph{International Conference on Machine Learning}.\hskip 1em plus 0.5em
  minus 0.4em\relax PMLR, 2016, pp. 1928--1937.

\bibitem{schulman2017proximal}
J.~Schulman, F.~Wolski, P.~Dhariwal, A.~Radford, and O.~Klimov, ``Proximal
  policy optimization algorithms,'' \emph{arXiv preprint arXiv:1707.06347},
  2017.

\bibitem{haarnoja2018soft}
T.~Haarnoja, A.~Zhou, P.~Abbeel, and S.~Levine, ``Soft actor-critic: Off-policy
  maximum entropy deep reinforcement learning with a stochastic actor,'' in
  \emph{International conference on machine learning}.\hskip 1em plus 0.5em
  minus 0.4em\relax PMLR, 2018, pp. 1861--1870.

\end{thebibliography}

\begin{IEEEbiography}[{\includegraphics[width=1in,height=1.25in,clip,keepaspectratio]{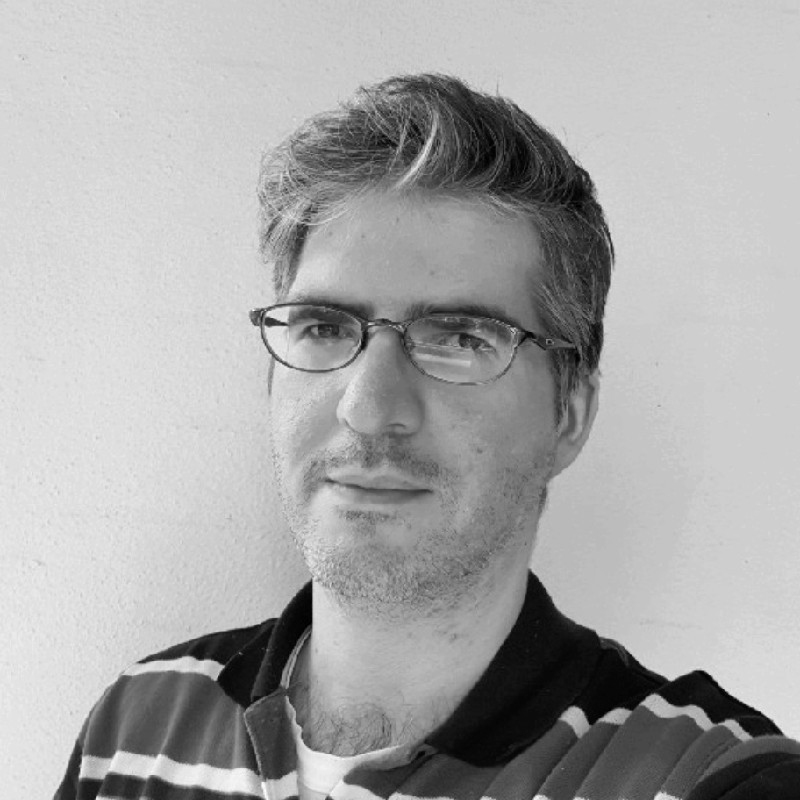}}]{Petros Giannakopoulos} is currently a PhD candidate at the Department of Informatics and Telecommunications, National and Kapodistrian University of Athens, Greece. His research is focused on deep reinforcement learning for audio applications. Previously, he earned an MSc in Computer Science and a BSc in Physics. He also has industry experience as a Machine Learning Engineer and as a Data Scientist, working for startups and research institutions.
\end{IEEEbiography}
\begin{IEEEbiography}[{\includegraphics[width=1in,height=1.25in,clip,keepaspectratio]{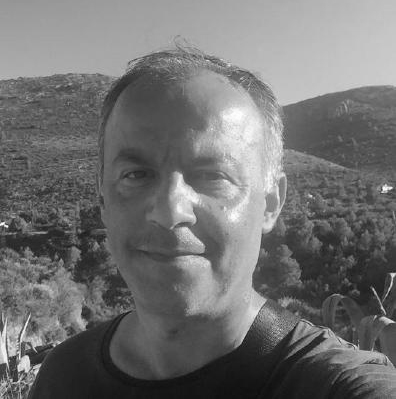}}]{Aggelos Pikrakis} received the Diploma Engineering degree from the University of Patras, in 1993, and the Ph.D. degree in Computer Science from the University of Athens, in 2002. He is currently an Assistant Professor at the Department of Informatics, University of Piraeus, Greece, teaching courses related to machine learning and audio processing. He is the co-inventor of AI/ML patents and he has coauthored two international textbooks in the English language and more than 50 refereed papers in international peer-reviewed scientific journals/conferences. His research interests include audio analysis, with an emphasis on machine learning algorithms, including deep neural networks, hidden Markov models, and Bayesian architectures and sequence alignment methods. His research work has received award recognition and he was a recipient of the 2019 EURASIP Meritorious Service Award.
\end{IEEEbiography}
\begin{IEEEbiography}[{\includegraphics[width=1in,height=1.25in,clip,keepaspectratio]{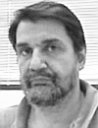}}]{Yannis Cotronis} is a Professor (retired) at the Department of Informatics and Telecommunications, National and Kapodistrian University of Athens. He holds a B.Sc. in Mathematics, M.Sc. and Ph.D. in Computing Science. His research interests include software engineering for parallel programming and e-science applications. He has taught a number of subjects in the core of Computer Science, his expert field being Parallel Systems and Programming. He has chaired EuroPVM-MPI 2001, Euromicro PDP 2000, 2011, 2016 and EuroMPI 2011.
\end{IEEEbiography}

\EOD
\end{document}